\documentclass[english]{iopart}
\usepackage[utf8]{inputenc}
\usepackage{array}
\usepackage{float}
\usepackage{graphicx}
\usepackage{iopams}

\makeatletter

\usepackage{setstack}

\makeatother

\usepackage{babel}
\begin{document}
\title{A quantum algorithm for finding collision-inducing disturbance vectors
in SHA-1}
\author{Jiheng Duan}
\address{Department of Physics and Chemistry, Faculty of Science and Technology,
University of Macau, Macau SAR, China}
\author{Minghui Li}
\address{Institute of Applied Physics and Materials Engineering, University
of Macau, Macau SAR, China}
\author{Hou Ian}
\address{Institute of Applied Physics and Materials Engineering, University
of Macau, Macau SAR, China}
\ead{houian@um.edu.mo}
\begin{abstract}
Modern cryptographic protocols rely on sophisticated hash functions
to generate quasi-unique numbers that serve as signatures for user
authentication and other security verifications. The security could
be compromised by finding texts hash-mappable to identical numbers,
forming so-called collision attack. Seeding a disturbance vector in
the hash mapping to obtain a successful collision is that a major
focus of cryptography study in the past two decades to improve hash
protocols. We propose an algorithm that takes advantage of entangled
quantum states for concurrent seeding of candidate disturbance vectors,
out of which the one entailing collision is selected through a combination
of quantum search, phase gating, diffusion gating, and information
feedbacks from classical computing machinery. The complexity reduction
is shown to be on the order of $\mathcal{O}(2^{n/2+1})$ where $n$
is the number of qubits encoding addresses. We demonstrate the practicality
of the proposed by an implementation scheme based on degenerate optical
parametric oscillators. 
\end{abstract}
\maketitle

\section{Introduction\label{sec:Introduction}}

Cryptographic hash functions appeared at the end of the twentieth
century and became one of the most significant concepts in key derivation,
password hashing, and digital signatures which were popularized since
1990s\ \cite{preneel2010first}. The most ubiquitous cryptographic
hash functions belong to the secure hash algorithm (SHA) family\ \cite{standard1995fips},
which includes the variants SHA-0 up to SHA-3. The SHA function converts
input messages into quasi-unique and orderless strings of a fixed
length. Since this length is much shorter than the lengths of original
messages, collisions are doomed to occur due to the pigeon-hole principle,
where two or more messages are converted to the same hash value. The
careful design of the SHA family, however, mandates that similar messages
lead to vastly distinct hash values, let alone messages of different
lengths. It was believed that hash collisions would never occur such
that authentication systems would be protected from fake signatures
from malware. The course took a sudden change in 2005 when the first
practical collision attack of SHA-1 was proposed by employing a differential
path method\ \cite{wang2005efficient,wang2005finding}. Since then,
more collision attacks have been found with relatively low algorithmic
complexities. Constructing differential paths essentially reduces
the time complexity of finding collision-inducing messages from brute-force
method and the introduction of disturbance vectors (DV) further simplifies
this construction~\cite{biham2005collisions,stevens2013new,szydlo2006collision,stevens2017speeding}.
Hence, finding the DVs becomes the key to finding the collision attacks
on SHA-1\ \cite{manuel2011classification}.

Despite the reduction in complexity, finding DV still remains a computationally
daunting task in a practice sense and often seeking a collision relies
on heuristic approaches. Quantum algorithms, nevertheless, poses unparalleled
advantage over classical algorithms in tackling specific computational
problems. The use of quantum computers would render some classically
improbable tasks tangible, granting the programmer quantum supremacy~\cite{arute2019quantum}.
Two particularly well-known problems falling in this category are
the factorization of large integers, solved by Shor's algorithm\ \cite{shor1994algorithms},
and the searching in unsorted list, solved by Grover's algorithm\ \cite{grover1996fast}.
The latter has been applied in numerous works to enhance computation
efficacy\ \cite{grassl2016applying,lavor2003grover,durr1996quantum}.
Therefore, it is only natural to take advantage of quantum parallelism
to improve the efficacy of seeking SHA-1 collisions but, to our knowledge,
scarcely have such quantum approaches been attempted. A quantum collision
search that elevates from classical semi-free-start collisions\ \cite{dobraunig2015analysis,mendel2013improving}
on reduced SHA-256 and SHA-512 was proposed last year\ \cite{hosoyamada2021quantum}.

Here, we propose a quantum algorithm for seeding DVs against SHA-1,
which combines diffusion and inversion gating from Grover's algorithm,
quantum phase gating, diffusion gating, and information feedback from
classical computing machinery, to greatly reduce the overall computation
complexity. The algorithm runs on a quantum machine of two $n=16\lambda$
qubit-long registers $R_{C}$ and $R_{W}$, where $\lambda$ is the
bit weight of the candidate DVs. While the control register $R_{C}$
stores the addresses of the DVs, the work register $R_{W}$ carries
the hash computation related information. The principles of the algorithm
apply to other SHA and MD5 hashing protocols as well, only that the
qubit length would vary. The concurrency of searching permitted by
the entanglement between $R_{1}$ and $R_{2}$ significantly reduces
the query complexity to $\mathcal{O}(2^{n/2})$, translating to a
total temporal complexity of $\mathcal{O}(2^{n/2+1})$. Beginning
with qubit initialization, the algorithm first uses Hadamard gates
to equally distribute probability among all DV addresses on $R_{1}$.
It is followed by phase gating, diffusion gating, and data for DV
validation to store corresponding entangled computation states on
$R_{2}$. Grover search then follows by performing the ``inversion
about average'' to elicit the address states with the greatest magnitude
of probability, which associate with valid DV states leading to collisions
in $R_{2}$. An algorithmic simulation based on pure-state quantum
evolution simulation package\ \cite{johansson2012qutip} verifies
the validity of the algorithm on $2^{7}$ space.

\begin{figure}
\includegraphics[trim={0 5cm 0 5cm},clip,width=10.5cm]{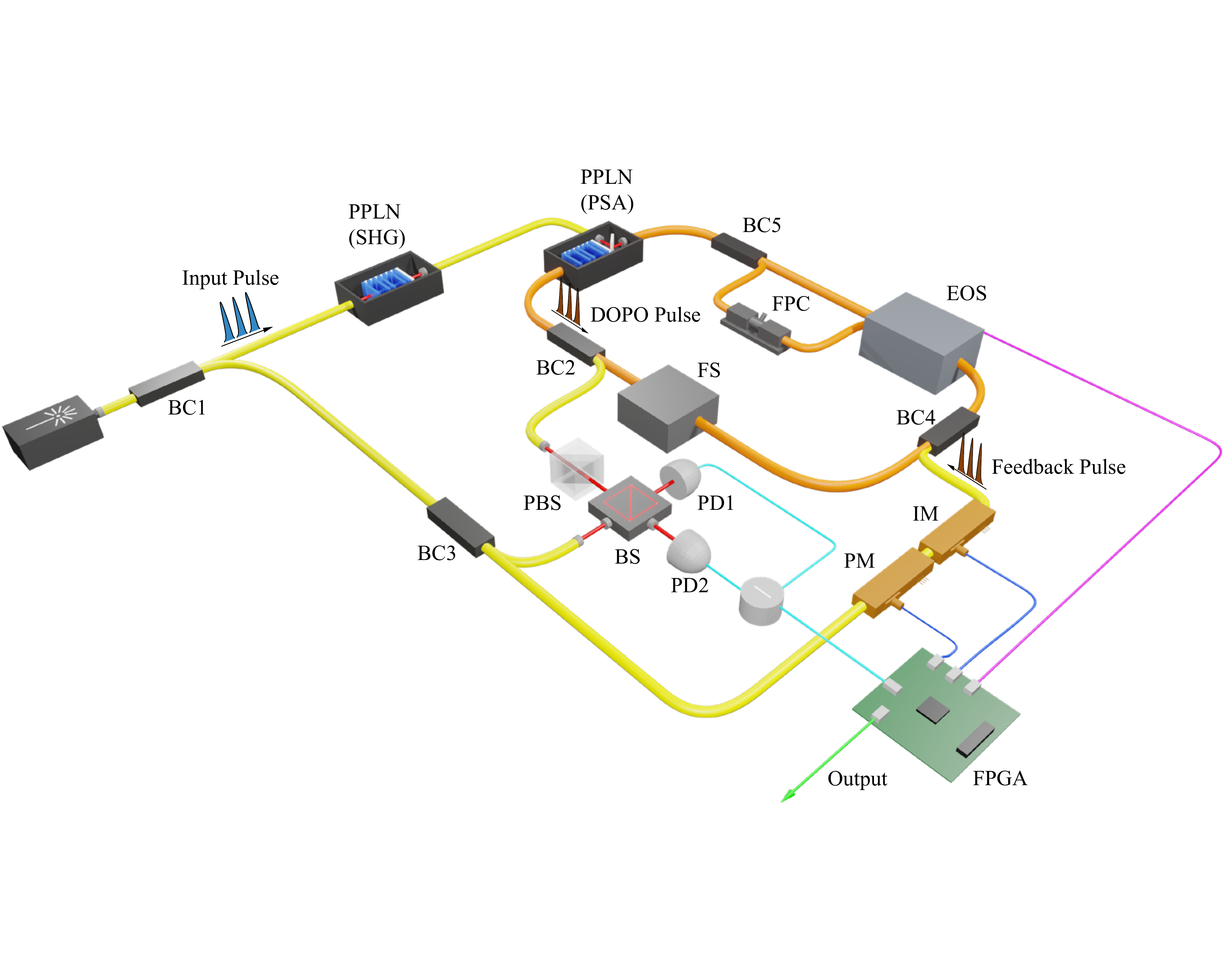}

\caption{Experimental schematic of a optical implementation of SHA-1 DV search
algorithm. FPGA: field-programmable gate array; PM: phase modulator;
IM: intensity modulator; EOS: electro-optical switch; FPC: fiber polarization
controller; FS: fiber stretcher; PD: photon diode; BS: beam splitter;
PBS: polarizing beam splitter.\label{fig:optial setup}}
\end{figure}

The proposed algorithm is optically implementable on degenerate optical
parametric oscillators (DOPOs). These parametric oscillators are coherent
laser pulses running in a ring optical cavity. The carrier phase and
the polarization direction of each such pulse are inseparable in a
sense equivalent to the inseparability of two quantum entangled states~\cite{you2021measurement}.
The coherence of these pulsed DOPOs can be maintained by phase-sensitive
amplification (PSA) in nonlinear optical crystals. The scale of pulse
width being as short as picosecond, a sufficiently long ring cavity
can store arbitrarily qubit-long information on a train of DOPOs,
where the entangled $R_{1}$ and $R_{2}$ data are encoded on the
phases and directions of the DOPOs, respectively. By virtue of their
coherent scalability, these special pulses have been employed to realize
graph-theoretic algorithms on the so-called coherent Ising machine\ \cite{inagaki2016coherent,marandi2014network,mcmahon2016fully}
as well as Shor's factoring algorithm~\cite{li2022scalable}. We
show here that all the quantum gating and feedback machinery involved
in DV searching described above are entirely implementable via optical
means and auxiliary classical computation. In other words, the proposed
algorithm is immediately accessible, where a viable experimental setup
is illustrated in Fig.\ \ref{fig:optial setup}. Before we describe
the setup in details in Sec.~\ref{sec:optical-implementation}, we
explain the SHA-1 hashing and its valid attacks in Sec.~\ref{sec:classical_coll},
which is followed by the descriptions of the quantum algorithm and
its simulated verification in Sec.~\ref{sec:Quantum-DV-seeding}.
The conclusions are given in Sec.~\ref{sec:Summary}. 

\section{Classical collision attacks of SHA-1\label{sec:classical_coll}}

The cryptographic hash function takes an input of a 512-bit block
and converts it into a 160-bit hash value output. Messages of lengths
less than $2^{64}$ bits long are first chopped into 512-bit blocks.
Following the Merkle-Damgard construction\ \cite{merkle1979secrecy},
each block is then sequentially fed into a compression loop, combining
with a 160-bit input chaining variable vector into a new 160-bit output
vector during, where the new vector will be used as the input for
compressing the next block in the next iteration. The chaining variables
are set initially to fixed values while the final chaining values
generated from compressing the last block forms the hash value, constituting
the encryption process of SHA-1. In the following, we simplify the
study by considering only one-block long messages.

Good cryptographic hash functions should be ``highly
sensitive'' to the messages, meaning that a small variation
in the message content results in a highly different hash output.
Collision attacks occur when two different messages generate the same
hash value. In other words, searching for collision attacks is to
find a message that shares the same hash value, while being distinct
from the intended message. Therefore, collision attack search serves
to validate the security of a hashing process. For SHA-1, its security
was bleached in 2005\ \cite{wang2005finding} when it was successfully
attacked through a collision realized by a well-constructed differential
path. In contrast to thorough brute-force search in the space of
all possible message inputs, the differential path construction is,
generally speaking, a bit-shifting process of the differences between
the values adopted by consecutive chaining variables. Uncovering the
differential path essentially reworks the chaining-variable dependent
compression loop such that a message with the same hash value can
be found without repetitively executing the compression process to
verify the hash output.

Therefore, the clever method has greatly reduced the time complexities
of finding collision attacks but constructing the differential path
is still computation intensive. To further mitigate the complexity,
one introduces disturbance vectors (DVs) defined as those difference
vectors that are reduced to zero after six compression iterations
between the colliding message and the original. In other words, such
DVs indicate the starting points of six-step local collisions, which
are characteristic of the SHA-1 hashing algorithm. One typical construction
of such local collisions was shown in Ref.~\cite{wang2005efficient},
where the chaining variables are reversely computed from the DVs that
lead to local collisions.

More specifically, a DV is a vector $v=(v_{0},v_{1},v_{2},\cdots,v_{79})$
comprising eighty 32-bit long words, i.e. each $v_{i}=(x_{0}^{i}x_{1}^{i}x_{2}^{i}\cdots x_{31}^{i})$
is a vector of binary numbers $x_{k}^{i}\in\{0,1\}$. These eighty
words are not entirely arbitrary but are correlated to each other
through the rule of message expansion amongst the 16-word partitions,
\begin{equation}
v_{i}=(v_{i-3}\oplus v_{i-8}\oplus v_{i-14}\oplus v_{i-16})\ll1,\label{eq:DV-expan-rel}
\end{equation}
where $\oplus$ denotes bitwise addition. In other words, the expansion
rule confines the space complexity for each $v$ to $\mathcal{O}(2^{16\times32})$.
For efficient collision search, only DVs with low Hamming weight $w$
in each component $v_{i}$ of $v$, i.e. those $v_{i}$ with a small
number of $x_{k}^{i}$ being one, are considered to constrain the
size of search space. Ref.~\cite{wang2005finding} use heuristics
to further confine their search space to reduce the search complexity:
only DVs with Hamming weight concentrated on the beginning and end
of each component, e.g. $x_{0}^{i}$, $x_{1}^{i}$, $x_{30}^{i}$,
and $x_{31}^{i}$ being one and the rest zero, are tested. The search
space is overall reduced to $\mathcal{O}(2^{n})$ where $n=16\lambda$,
and $\lambda$ is the bit weight describes the range of $k$ that
$x_{k}^{i}$ could be $1$. A recent analysis of cost functions has
further improved the space complexity to $65\times{512\choose w}$
where $w$ is the Hamming weight\ \cite{manuel2011classification}.

It can be noticed from Eq.~(\ref{eq:DV-expan-rel}) that the bits
of a DV are not independent; rather, the second and subsequent words
of each DV are dependent on the first word. Therefore, if the valid
DVs are juxtaposed as columns where each row denotes a word of certain
Hamming weight, they will be tabulated in a matrix where the appearances
of their first words will depart from each other by a few rows\ \cite{manuel2011classification}.
According to such a scheme of tabulation, valid DVs can be classified
into two types. Searching for the locations of these first word appearances
would be the goal of a practical setup discussed in Sec.~\ref{sec:sp-type-I-DV}. 

\section{Quantum algorithm for seeding disturbance vector\label{sec:Quantum-DV-seeding}}

With the advent of quantum computer, the so-called quantum supremacy
over classical algorithms is achieved when the unique features of
quantum systems, such as superpositions and entangled states, are
taken advantage of to obtain computing parallelism not available to
classical computing machinery. From the perspective of complexity,
this parallelism is rendered as a reduction across different complexity
classes, often from the exponential class to the polynomial class.
We see from the last section that classical collision search, though
having already its complexity reduced through the assistance of heuristics,
remains a hard problem due to the vast search space. In the following,
we present our quantum search algorithm for a DV seed out of which
would grow a collision in SHA-1 hash value.

As we discussed in the last section, characterizing a DV requires
the knowledge of any sixteen consecutive components $(v_{i},\cdots,v_{i+15})$
by employing the forward (or backward) message expansion formula.
In particular, we can characterize all possible DVs by exhausting
on the first sixteen components $(v_{0},\cdots,v_{15})$. Following
the classical studies of SHA-1 collision searching, we assume a low
bit weight $\lambda$ for the occurrences of valid DVs such that the
candidate DVs can be expressed as the concatenation $\mu=X^{0}X^{1}\cdots X^{15}$
of 16 $\lambda$-bit long strings $X^{i}=x_{0}^{i}x_{1}^{i}\cdots x_{\lambda-1}^{i}$.
We adopt the convention that the leftest bit $x_{0}^{0}$ to be the
least significant bit. For a hardware perspective, the candidate DVs
are stored in a $n$-qubit quantum work register $R_{W}$, whose address
is specified by another $n$-qubit quantum control register $R_{C}$
it entangles with. An eigenvector $|\psi_{C},\psi_{W}\rangle$ then
denotes an associated pair of address data and content data for each
candidate DV. The concurrency of the quantum algorithm is thus reflected
by the possibility of searching for collision-inducing DVs in a linear
superposition of these eigenvectors.

\begin{figure}[H]
\includegraphics[trim={0 0 0 0},clip,width=10.5cm]{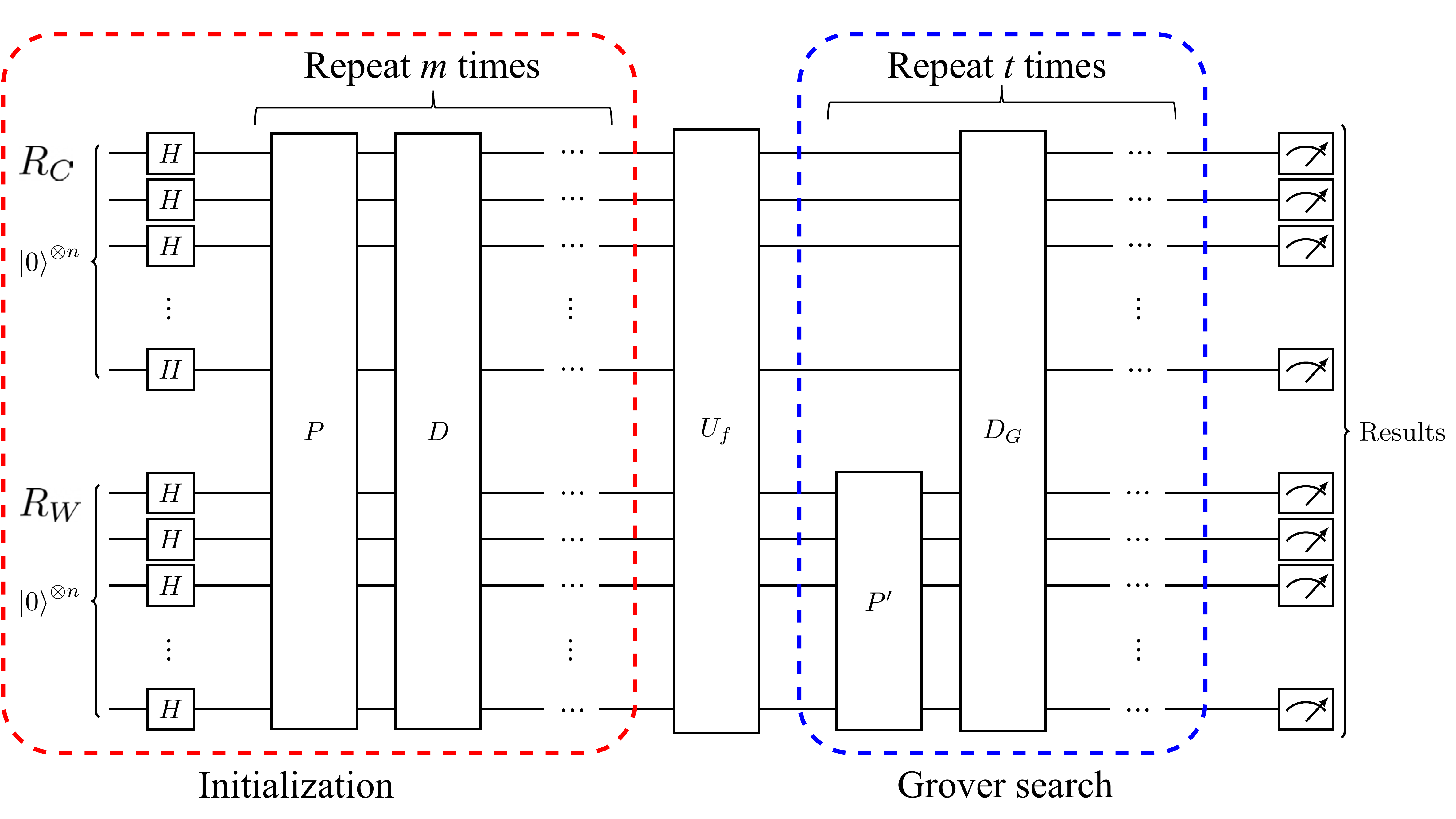}

\caption{An implementation of the DV-search algorithm by a quantum circuit.
Starting with $n$ qubits with each state $|0\rangle$ in the first
and work registers. The number of repeat times of Grover search $t=\mathcal{O}(2^{n/2})$.
Initialization and Grover search will operate on these direct product
states which turn the qubits system representing an entangled state
containing marked states with larger norms of their coefficients.
Then, measuring all the qubits and readouts will give the information
of valid DVs.\label{fig:quantum-circuit}}
\end{figure}

In other words, for the SHA-1 hashing, both $R_{C}$ and $R_{W}$
registers are $n=16\lambda$ qubit long, rendering the DV search space
$S=\{\zeta|0\le\zeta\le2^{n}-1\}$. Encoding the classical data of
$S$ into the Hilbert space, each of the two registers states are
distributed with equal probability using the Hadamard gates before
we could proceed to the next operation, as shown in Fig~\ref{fig:quantum-circuit}.
To generate the entanglement between $R_{C}$ and $R_{W}$ such that
the DV data registered in $R_{W}$ is directly correlated with the
address in $R_{C}$, a set of phase gates $P$ and diffusion gates
$D$ (shown as the red dashed box of Fig.~\ref{fig:quantum-circuit})
is applied. That is, the joint eigenstates of the quantum computer
would have the address value record the original value of candidate
DVs so that subsequent operations on $R_{W}$ would leave the register
$R_{C}$ unchanged and the $R_{C}$ states that remain would contain
the value of valid DVs associated with hashing collisions. 

Conventionally, the correlation between $R_{C}$ and $R_{W}$ is formed
through entanglement operations implemented using elementary logical
gates such as $\sqrt{\mathrm{iSWAP}}$ gates based on the superexchange
(SE) interaction, which were used for generating Einstein-Podolsky-Rosen
(EPR) pairs among ultracold Rb atoms~\cite{trotzky2008time} and
Greenberger-Horne-Zeilinger (GHZ) state~\cite{song201710}. The SE
interaction permits long-range correlation between two configurations
of two spin particles inside a double-well potential. However, such
ubiquitous methods are not suitable for the scenario here because
of the extremely sophisticated combinations of quantum logical gates.
The phase and diffusion gates in our approach realize approximately
the same entangled state except for a tail part $\left|\psi\right\rangle $
of negligible probability without heavily taxing the complexity of
the entire algorithm. Once the correlation is generated, a DV-validity
authenticating gate $U_{f}$ is applied to change the data in the
work register $R_{W}$ to the hashed value: $|\Psi_{f}\rangle=U_{f}\left|\Psi_{0}\right\rangle $
corresponding to the entangled data in the control register $R_{C}$.
Finally, a Grover search is executed to select the states with the
colliding hashed value. In the subsections below, we detail the quantum
operations needed to implement the multiple step introduced above.

\subsection{State preparations on the registers\label{subsec:state_prep}}

Before the machine can correctly select the valid DVs from the control
register, it is prepared with an initialization process that converts
the preliminary zero state $|0\rangle$ of the $2n$ qubits into a
superposition of equally weighted candidate states, using Hadamard
gates:
\begin{equation}
|\Psi_{0}\rangle=\frac{1}{\sqrt{N}}\sum_{\alpha=0}^{\sqrt{N}-1}\sum_{\beta=0}^{\sqrt{N}-1}|\alpha,\beta\rangle,\label{eq:initial_state}
\end{equation}
where $N=2^{2n}$ is the total number of possible combinations of
$R_{C}$ and $R_{W}$ data. Since candidate DVs serve both as the
address in $R_{C}$ and the work content in $R_{W}$, only the eigenstates
with identical $R_{C}$ and $R_{W}$ data are wanted. The phase gates
$P$ and the diffusion gates $D$, indicated by the red dashed square
in Fig~\ref{fig:quantum-circuit}, are therefore used to filter out
the undesired eigenstates to retain the selection

\begin{equation}
|\Psi\rangle=\frac{1}{N^{1/4}}\sum_{\zeta=0}^{\sqrt{N}-1}|\zeta,\zeta\rangle,\label{eq:desired _state}
\end{equation}
The undesired states constitute a tail state $\left|\psi\right\rangle $,
which would contribute a negligible probability after multiple iterations
of $DP$ gating, i.e. after $m$ iterations on the initial state,
we have

\begin{equation}
(DP)^{m}|\Psi_{0}\rangle=c|\Psi\rangle+c'|\psi\rangle,\label{eq:state_after_DP}
\end{equation}
where $c,c'\in\mathbb{R}$ and $|c|^{2}\gg|c'|^{2}$. To show this,
we can refer to the proof steps of the effectiveness of the Grover
search~\cite{grover1996fast} by treating the phase and diffusion
gating here as inversion about average and diffusion of probabilities
among all eigenstates, respectively. In other words, the operation
is a generalized Grover search that amplifies the probabilities of
multiple eigenstates instead of a single one, at the end of which
the remaining states are the ones that have formed entanglement between
$\left|\zeta\right\rangle $ in $R_{C}$ and $\left|\zeta\right\rangle $
in $R_{W}$, i.e. 

\begin{equation}
(DP)^{m}|\Psi_{0}\rangle\approx|\Psi\rangle.\label{eq:approx_state_after_DP}
\end{equation}

The phase and diffusion gating is readily decomposable into elementary
single- and double-qubit gates~\cite{barenco1995elementary,shende2006synthesis}.
Their full $N\times N$ matrix representations are 

\begin{equation}
[P]_{ij}=\left\{\begin{array}{r@{\quad}cr}
0  & i\neq j\\
-1 & i=j=\sqrt{N}\zeta+\zeta+1\\
1  &\mathrm{otherwise}
\end{array}\right.
\hfill[D]_{ij}=\left\{\begin{array}{r@{\quad}cr}
\frac{2}{N} & i\neq j\\
\frac{2}{N}-1 & i=j
\end{array},\right.\label{eq:f matix_elem_inver}
\end{equation}
where $\zeta$ ranges over $0$ to $\sqrt{N}-1$. The decomposition
of the phase gate can be accomplished in two step: (i) decompose $P$
into the product $P=\Pi_{\nu=0}^{\sqrt{N}-1}\Lambda_{\nu}$ where
the elements of the $\Lambda_{\nu}$ matrices are

\begin{equation}
[\Lambda_{\nu}]_{ij}=\left\{\begin{array}{r@{\quad}cr}
0 & i\neq j\\
-1 & i=j=\sqrt{N}\nu+\nu+1;\\
1 & \mathrm{otherwise}
\end{array}\right.\label{eq:phase-gate-decomp}
\end{equation}
(ii) decompose these $\Lambda^{\nu}$ into elementary controlled gates
such as CZ and $Z$-rotation gates~\cite{mandviwalla2018implementing}.
To decompose the diffusion gate $D$, (i) first rewrite it in terms
of the outer product of the initial machine state: $D=2|\Psi_{0}\rangle\langle\Psi_{0}|-I$;
(ii) then using the tensor product $H^{\otimes2n}$ of Hadamard gates
$H$, one can express $D=H^{\otimes2n}\left(2|0\rangle\langle0|-I\right)H^{\otimes2n}$
where the middle factor is just a $2n$-qubit CZ gate. The latter
is again further decomposable into double-qubit CZ gates and $Z$-rotation
gates.

After the entanglement, a DV-validity authenticating gate $U_{f}$
is operated on the machine state to switch the eigenstates of the
work register $R_{W}$ to its hashed value, i.e.

\begin{equation}
U_{f}|\zeta,\zeta\rangle=|\zeta,f(\zeta)\rangle,\label{eq:DV-validity auth}
\end{equation}
where the SHA-1 authenticating function $f:\mathbb{N}\times\mathbb{N}\rightarrow\mathbb{N}$
assigns a zero value if $\zeta$ is a collision-inducing DV or otherwise
a non-zero value $\mu_{\zeta}$. For example, if $\xi$ is such a
valid DV, $U_{f}\left|\xi,\xi\right\rangle =\left|\xi,0\right\rangle $.
When $U_{f}$ operates on all eigenstates, the final machine state
has the eigenstates partitioned into two groups: 

\begin{equation}
|\Psi_{f}\rangle\approx\frac{1}{N^{1/4}}\left(\sum_{\zeta\in V}|\zeta,0\rangle+\sum_{\zeta\notin V}|\zeta,\mu_{\zeta}\rangle\right),\label{eq:final state}
\end{equation}
where $V$ indicates the set of all valid DVs. 

\subsection{Grover search\label{subsec:Grover-search}}

The second major part of the algorithm is a standard Grover search,
indicated by the blue dashed square in Fig~\ref{fig:quantum-circuit},
which is also implementable using single- and double-qubit gates.
The Grover search performed on the work register amplifies the probabilities
of the states carrying specific data, which in our case is the zero
value for valid DVs. In order to search for the zero states of $R_{W}$,
it is necessary to redefine the inversion gate $P_{G}$ as the tensor
product $I\otimes P'$, where the phase gate $P'$ is a $\sqrt{N}\times\sqrt{N}$
matrix with elements 
\begin{equation}
[P']_{ij}=\left\{\begin{array}{r@{\quad}cr}
0 & i\neq j\\
-1 & i=j=1\\
1 & \mathrm{otherwise}
\end{array},\right.\label{eq:f inversion_matrxi_ele}
\end{equation}
on the work register $R_{W}$. The diffusion gate 
\begin{equation}
D_{G}=2|\Psi_{f}\rangle\langle\Psi_{f}\vert-I,\label{eq:def_inversion_diffu}
\end{equation}
though similar to the one in last section, is defined with $\left|\Psi_{f}\right\rangle $
instead of $\left|\Psi_{0}\right\rangle $. After $N^{1/4}$ iterations
of $D_{G}P_{G}$ , the probabilities of the states with valid DVs
are amplified.

\begin{figure}
\includegraphics[trim={5cm 18cm 30cm 0},clip,width=10.5cm]{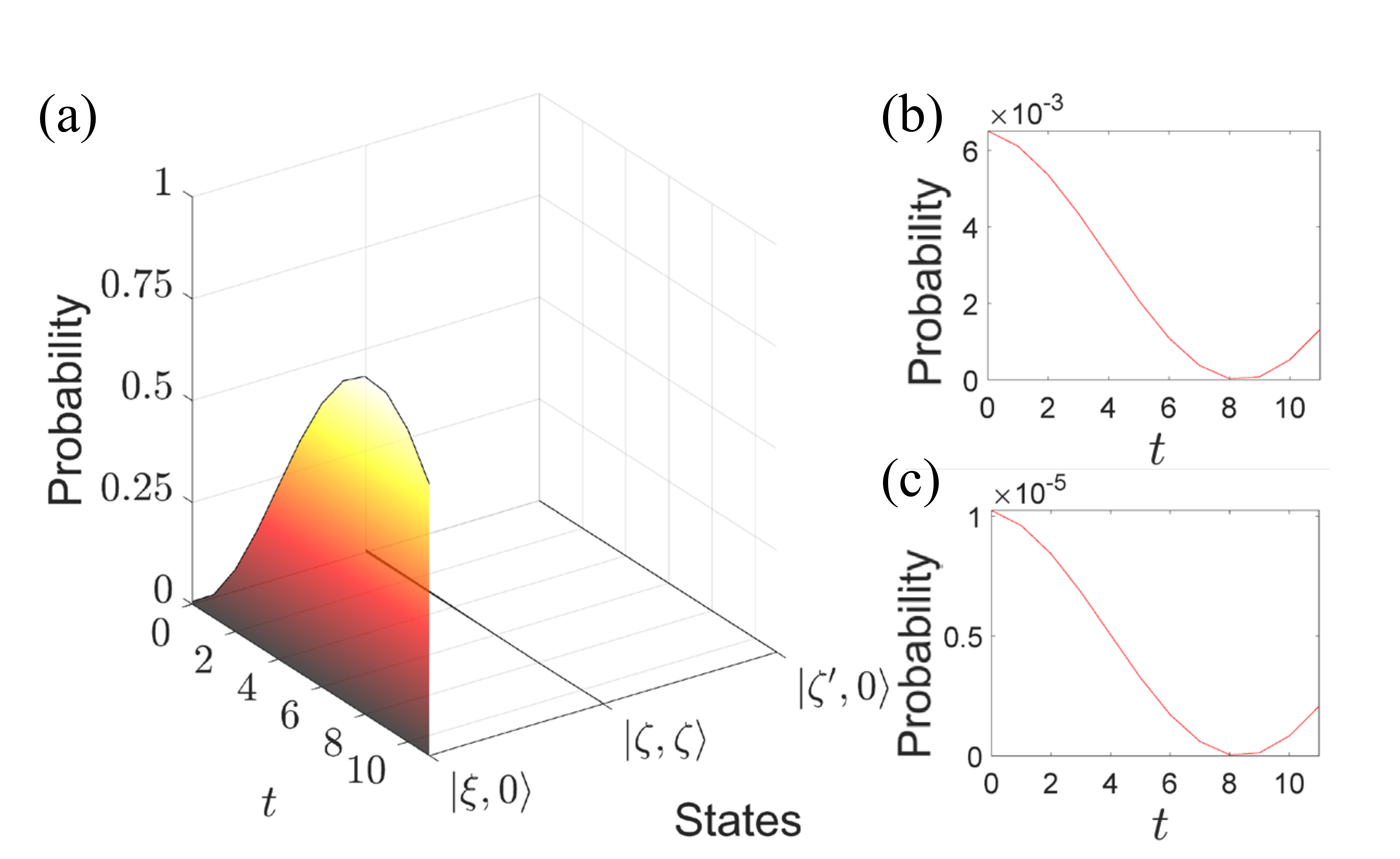}

\caption{Probability variations of typical eigenstates of a simulation of the
collision-inducing DV searching algorithm with $2n=14$ qubits, along
with the number $t\in\{1,2,\dots,\lfloor2^{n/2}\rfloor\}$ of Grover
search iterations (applications of $D_{G}P_{G}$ gates) where both
the control register $R_{C}$ and the work register $R_{W}$ have
$n=7$ qubits, using a quantum evolution simulation package~\cite{johansson2012qutip}.
In (a), $|\xi,0\rangle$ indicates the states with collision-inducing
DV $\xi$ which only equals zero in this simulation; $|\zeta,\zeta\rangle$
the other states with non-zero $\zeta$; $\left|\zeta',0\right\rangle $
the tail states with diminishing probability, although having $R_{W}$
equals zero. The results show that the algorithm sufficiently amplifies
the probability of the marked states $|\xi,0\rangle$. (b) and (c)
shows the magnified views of the probabilities of $|\zeta,\zeta\rangle$
and $\left|\zeta',0\right\rangle $, respectively.~\label{fig:algorithm-simulation}}
\end{figure}

We have run a simulation of the collision-inducing DV searching algorithm
with $2n=14$ qubits using a quantum evolution simulation package~\cite{johansson2012qutip}.
The machine states are initialized at $|0\rangle$ and $m=\lfloor2^{n/2}\rfloor$
iterations of $DP$ gating are applied. Figure\ \ref{fig:algorithm-simulation}
shows the probabilities of the three typical eigenstates $|\xi,0\rangle$,
$|\zeta,\zeta\rangle$, and $|\zeta',0\rangle$ from Eq.~(\ref{eq:final state})
at every iteration of $D_{G}P_{G}$ gating during the Grover search.
The eigenstate $\left|\xi,0\right\rangle $ represents the case of
valid DV $\xi$ that would induce a hash collision, i.e. $f(\xi)=0$;
the eigenstate $\left|\zeta,\mu_{\zeta}\right\rangle $ where $\zeta$
is not one of those valid $\xi$ represents the case of non-valid
DV; the eigenstate $\left|\zeta',0\right\rangle $ represents the
tail state case where $f(\zeta')=0$ even though $\zeta'$ is not
a valid DV. Since the authenticating $U_{f}$ is conducted outside
of the quantum circuit, we have used $\left|0,0\right\rangle $ for
the $\left|\xi,0\right\rangle $ case and $\left|\zeta,\zeta\right\rangle $
for the $\left|\zeta,\mu_{\zeta}\right\rangle $ case without loss
of generality in the quantum simulation. 

One can observe that before the Grover search begins, the tail state
$\left|\psi\right\rangle $ which comprises the $\left|\zeta',0\right\rangle $
eigenstates has the much suppressed probability $1.024\times10^{-5}$
compared to the probability $6.512\times10^{-3}$ of the valid-DV
states and the regular non-valid-DV states. After Grover search begins,
each iteration of $D_{G}P_{G}$ gating further suppress the probability
of $\left|\psi\right\rangle $ to a negligible value; the magnified
view of this probability is given in Fig.~\ref{fig:algorithm-simulation}(c).
Meanwhile, the state $\left|\xi,0\right\rangle $ has its probability
be amplified to a significant value while the state $\left|\zeta,\mu_{\zeta}\right\rangle $
experiences a great probability reduction (magnified view shown in
Fig.~\ref{fig:algorithm-simulation}(b)) during the $D_{G}P_{G}$
gatings, although both of them start off from the same probability
value. Therefore, the distinguishing ability of the valid disturbance
vector from the non-valid in the proposed algorithm is verified.

\subsection{Complexity estimation\label{subsec:Complexity-estimation}}

According to classical computation theory, $\mathcal{O}(2^{\kappa})$
classical queries are required to compute all possible results encoded
by $\kappa$ classical bits in the worst case scenario. In contrast,
the complexity metric that measures the DV-seeding quantum algorithm
is contributed by the phase gating, the diffusion gating, and the
Grover search that are characterized by a black box function, known
as the query model\cite{ambainis2000quantum,ambainis2006polynomial,bernstein1997quantum}.
The initialization using Hadamard gates and the following DV-authenticating
operator $U_{f}$, whose complexity can be regarded as $\mathcal{O}(1)$,
are excluded from the query searching part\ \cite{durr1996quantum}.
The phase gate $P$ and diffusion gate $D$ are admitted to have a
query time $m\approx2^{n/2}$ which contributes a complexity $\mathcal{O}(2^{n/2})$
of the operations $DP$. The Grover search is applied on the work
register $R_{W}$, which contributes a query complexity of $\mathcal{O}(2^{n/2})$
under the query time $t=2^{n/2}$. Under the implementation of a quantum
circuit, the number of quantum gates that implement the black-box
operators is polynomial, rather than exponential, in the number of
input qubits. Since the \emph{DP} step and the Grover search are sequential,
the total query complexity of the algorithm is $\mathcal{O}(2^{n/2+1})$
in the search space $S=\{\zeta|0\le\zeta\le2^{n}-1\}$, which is vastly
sped up from the $\mathcal{O}(2^{n})$ queries for classical complexity.

\section{Specification on type-I DV\label{sec:sp-type-I-DV}}

Current cryptographic studies have categorized the occurrences of
valid DVs into two types, according to their clustering in certain
subsets of the entire search space. Then given these category information,
the quantum search of valid DVs can be confined to the subsets, which
greatly reduces the search complexity. Specifically, the address of
the first word of the type-I DVs, for instance, only appears in a
$2^{5}$space out of $2^{512}$\ \cite{manuel2011classification}.
Our quantum algorithm described in last section can still apply to
this scenario of restricted search space. Moreover, under the consideration
of the limited quantum resources, the confinement of the search space
significantly reduces the quantum resources, such as the number of
qubits and elementary quantum gates employed in the quantum algorithm,
by increasing the classical restriction.

All type-I DVs are contained in a table constructed from a set of
consecutive 16-bit words $\{v_{i}\}$, which satisfies the message
expansion relation Eq.\ \ref{eq:DV-expan-rel}, for each $v_{i}$
represents a line. DVs in this type can be mapped into this table,
operated by their unique bit-shift, starting from different lines.
The table can be infinitely expanded forward or backward but is manually
limited as it is sufficient to cover all type-I DVs found by previous
works. Notice that it is possible to expand the rest part of a type-I
DV after knowing its starting line, which can be used to further reduce
the search space. In this case, the search space is shrunk from $S$
into a deterministic table shown in Tab.\ \ref{tab:Eigenstates from type-I DV-sheet},
and the number of qubits in the control and work register are both
$n=5$. The initial state is shown in this form

\begin{equation}
|\Psi_{0}\rangle=\frac{1}{2^{5}}\sum_{\alpha=0}^{2^{5}-1}\sum_{\beta=0}^{2^{5}-1}|\alpha,\beta\rangle=\frac{1}{2^{5}}\left(|\Psi_{T}\rangle+|\psi\rangle\right),\label{eq:partial example of out algorithm}
\end{equation}
where $|\Psi_{T}\rangle$ is the summation of eigenstates $|l,u_{l}\rangle$
listed in Tab.\ \ref{tab:Eigenstates from type-I DV-sheet} and tail
state $|\psi\rangle$. After applying the phase gates, diffusion gates,
and DV-validity authenticating gate, the machine state $|\Psi_{f}\rangle$
intends to contain these marked eigenstates $|l,0\rangle$ for $l\in\left\{ 0,1,2,3,7,11,13,17\right\} $,
with the probability of the tail state being diminished to zero. Then,
the Grover search is acted on the system and amplifies the magnitudes
of the marked eigenstates, which are the states expected to be measured.

\begin{table}[H]
\textbf{\caption{\label{tab:Eigenstates from type-I DV-sheet}The eigenstates $|l,u_{l}\rangle$
inside $|\Psi_{T}\rangle$ based on the table of the expanded type-I
DV.}
}{\scriptsize{}}%
\begin{tabular}{c|cc|cc|c}
\hline 
The first 18 lines of the &  &  & Starting points  &  & \tabularnewline
Type-I & Order & $|l\rangle\in R_{C}$ & of 6-step & $|u_{l}\rangle\in R_{W}$ & $|l,u_{l}\rangle$ \tabularnewline
Disturbance vector&  &  & local collisions &  & \tabularnewline
\hline 
 &  &  &  &  & \tabularnewline
{\scriptsize{}-{}-{}-{}-{}-{}-{}-{}-{}-{}-{}-{}-{}-{}-{}-{}-{}-{}-{}-{}-{}-{}-{}-{}-{}-{}-{}-{}-{}-{}-{}-{}-} & {\scriptsize{}1} & {\scriptsize{}$|0\rangle$} & {\scriptsize{}0} & {\scriptsize{}$|0\rangle$} & {\scriptsize{}$|0,0\rangle$}\tabularnewline
{\scriptsize{}-oo-{}-{}-{}-{}-{}-{}-{}-{}-{}-{}-{}-{}-{}-{}-{}-{}-{}-{}-{}-{}-{}-{}-{}-{}-{}-{}-{}-{}-} & {\scriptsize{}2} & {\scriptsize{}$|1\rangle$} & {\scriptsize{}6} & {\scriptsize{}$|6\rangle$} & {\scriptsize{}$|1,6\rangle$}\tabularnewline
{\scriptsize{}-{}-{}-{}-{}-{}-{}-{}-{}-{}-{}-{}-{}-{}-{}-{}-{}-{}-{}-{}-{}-{}-{}-{}-{}-{}-{}-{}-{}-{}-{}-{}-} & {\scriptsize{}3} & {\scriptsize{}$|2\rangle$} & {\scriptsize{}0} & {\scriptsize{}$|0\rangle$} & {\scriptsize{}$|2,0\rangle$}\tabularnewline
{\scriptsize{}-o-o-{}-{}-{}-{}-{}-{}-{}-{}-{}-{}-{}-{}-{}-{}-{}-{}-{}-{}-{}-{}-{}-{}-{}-{}-{}-{}-{}-} & {\scriptsize{}4} & {\scriptsize{}$|3\rangle$} & {\scriptsize{}10} & {\scriptsize{}$|10\rangle$} & {\scriptsize{}$|3,10\rangle$}\tabularnewline
{\scriptsize{}o-{}-{}-{}-{}-{}-{}-{}-{}-{}-{}-{}-{}-{}-{}-{}-{}-{}-{}-{}-{}-{}-{}-{}-{}-{}-{}-{}-{}-{}-{}-} & {\scriptsize{}5} & {\scriptsize{}$|4\rangle$} & {\scriptsize{}1} & {\scriptsize{}$|1\rangle$} & {\scriptsize{}$|4,1\rangle$}\tabularnewline
{\scriptsize{}o-{}-{}-{}-{}-{}-{}-{}-{}-{}-{}-{}-{}-{}-{}-{}-{}-{}-{}-{}-{}-{}-{}-{}-{}-{}-{}-{}-{}-{}-{}-} & {\scriptsize{}6} & {\scriptsize{}$|5\rangle$} & {\scriptsize{}1} & {\scriptsize{}$|1\rangle$} & {\scriptsize{}$|5,1\rangle$}\tabularnewline
{\scriptsize{}o-o-{}-{}-{}-{}-{}-{}-{}-{}-{}-{}-{}-{}-{}-{}-{}-{}-{}-{}-{}-{}-{}-{}-{}-{}-{}-{}-{}-{}-} & {\scriptsize{}7} & {\scriptsize{}$|6\rangle$} & {\scriptsize{}5} & {\scriptsize{}$|5\rangle$} & {\scriptsize{}$|6,5\rangle$}\tabularnewline
{\scriptsize{}-o-{}-{}-{}-{}-{}-{}-{}-{}-{}-{}-{}-{}-{}-{}-{}-{}-{}-{}-{}-{}-{}-{}-{}-{}-{}-{}-{}-{}-{}-} & {\scriptsize{}8} & {\scriptsize{}$|7\rangle$} & {\scriptsize{}2} & {\scriptsize{}$|2\rangle$} & {\scriptsize{}$|7,2\rangle$}\tabularnewline
{\scriptsize{}-{}-{}-{}-{}-{}-{}-{}-{}-{}-{}-{}-{}-{}-{}-{}-{}-{}-{}-{}-{}-{}-{}-{}-{}-{}-{}-{}-{}-{}-{}-{}-} & {\scriptsize{}9} & {\scriptsize{}$|8\rangle$} & {\scriptsize{}0} & {\scriptsize{}$|0\rangle$} & {\scriptsize{}$|8,0\rangle$}\tabularnewline
{\scriptsize{}-oo-{}-{}-{}-{}-{}-{}-{}-{}-{}-{}-{}-{}-{}-{}-{}-{}-{}-{}-{}-{}-{}-{}-{}-{}-{}-{}-{}-{}-} & {\scriptsize{}10} & {\scriptsize{}$|9\rangle$} & {\scriptsize{}6} & {\scriptsize{}$|6\rangle$} & {\scriptsize{}$|9,6\rangle$}\tabularnewline
{\scriptsize{}o-{}-{}-{}-{}-{}-{}-{}-{}-{}-{}-{}-{}-{}-{}-{}-{}-{}-{}-{}-{}-{}-{}-{}-{}-{}-{}-{}-{}-{}-{}-} & {\scriptsize{}11} & {\scriptsize{}$|10\rangle$} & {\scriptsize{}1} & {\scriptsize{}$|1\rangle$} & {\scriptsize{}$|10,1\rangle$}\tabularnewline
{\scriptsize{}o-{}-{}-{}-{}-{}-{}-{}-{}-{}-{}-{}-{}-{}-{}-{}-{}-{}-{}-{}-{}-{}-{}-{}-{}-{}-{}-{}-{}-{}-{}-} & {\scriptsize{}12} & {\scriptsize{}$|11\rangle$} & {\scriptsize{}1} & {\scriptsize{}$|1\rangle$} & {\scriptsize{}$|11,1\rangle$}\tabularnewline
{\scriptsize{}o-{}-{}-{}-{}-{}-{}-{}-{}-{}-{}-{}-{}-{}-{}-{}-{}-{}-{}-{}-{}-{}-{}-{}-{}-{}-{}-{}-{}-{}-{}-} & {\scriptsize{}13} & {\scriptsize{}$|12\rangle$} & {\scriptsize{}1} & {\scriptsize{}$|1\rangle$} & {\scriptsize{}$|12,1\rangle$}\tabularnewline
{\scriptsize{}-{}-{}-{}-{}-{}-{}-{}-{}-{}-{}-{}-{}-{}-{}-{}-{}-{}-{}-{}-{}-{}-{}-{}-{}-{}-{}-{}-{}-{}-{}-{}-} & {\scriptsize{}14} & {\scriptsize{}$|13\rangle$} & {\scriptsize{}0} & {\scriptsize{}$|0\rangle$} & {\scriptsize{}$|13,0\rangle$}\tabularnewline
{\scriptsize{}-{}-{}-{}-{}-{}-{}-{}-{}-{}-{}-{}-{}-{}-{}-{}-{}-{}-{}-{}-{}-{}-{}-{}-{}-{}-{}-{}-{}-{}-{}-{}-} & {\scriptsize{}15} & {\scriptsize{}$|14\rangle$} & {\scriptsize{}0} & {\scriptsize{}$|0\rangle$} & {\scriptsize{}$|14,0\rangle$}\tabularnewline
{\scriptsize{}-o-{}-{}-{}-{}-{}-{}-{}-{}-{}-{}-{}-{}-{}-{}-{}-{}-{}-{}-{}-{}-{}-{}-{}-{}-{}-{}-{}-{}-{}-} & {\scriptsize{}16} & {\scriptsize{}$|15\rangle$} & {\scriptsize{}2} & {\scriptsize{}$|2\rangle$} & {\scriptsize{}$|15,2\rangle$}\tabularnewline
{\scriptsize{}-{}-{}-{}-{}-{}-{}-{}-{}-{}-{}-{}-{}-{}-{}-{}-{}-{}-{}-{}-{}-{}-{}-{}-{}-{}-{}-{}-{}-{}-{}-{}-} & {\scriptsize{}17} & {\scriptsize{}$|16\rangle$} & {\scriptsize{}0} & {\scriptsize{}$|0\rangle$} & {\scriptsize{}$|16,0\rangle$}\tabularnewline
{\scriptsize{}o-o-{}-{}-{}-{}-{}-{}-{}-{}-{}-{}-{}-{}-{}-{}-{}-{}-{}-{}-{}-{}-{}-{}-{}-{}-{}-{}-{}-{}-} & {\scriptsize{}18} & {\scriptsize{}$|17\rangle$} & {\scriptsize{}5} & {\scriptsize{}$|5\rangle$} & {\scriptsize{}$|17,5\rangle$}\tabularnewline
 &  &  &  &  & \tabularnewline
\hline 
\end{tabular}
\end{table}

\section{Implementation using DOPO network\label{sec:optical-implementation}}

\begin{figure}
    \includegraphics[trim={2cm 8cm 0cm 8cm},clip,width=10.5cm]{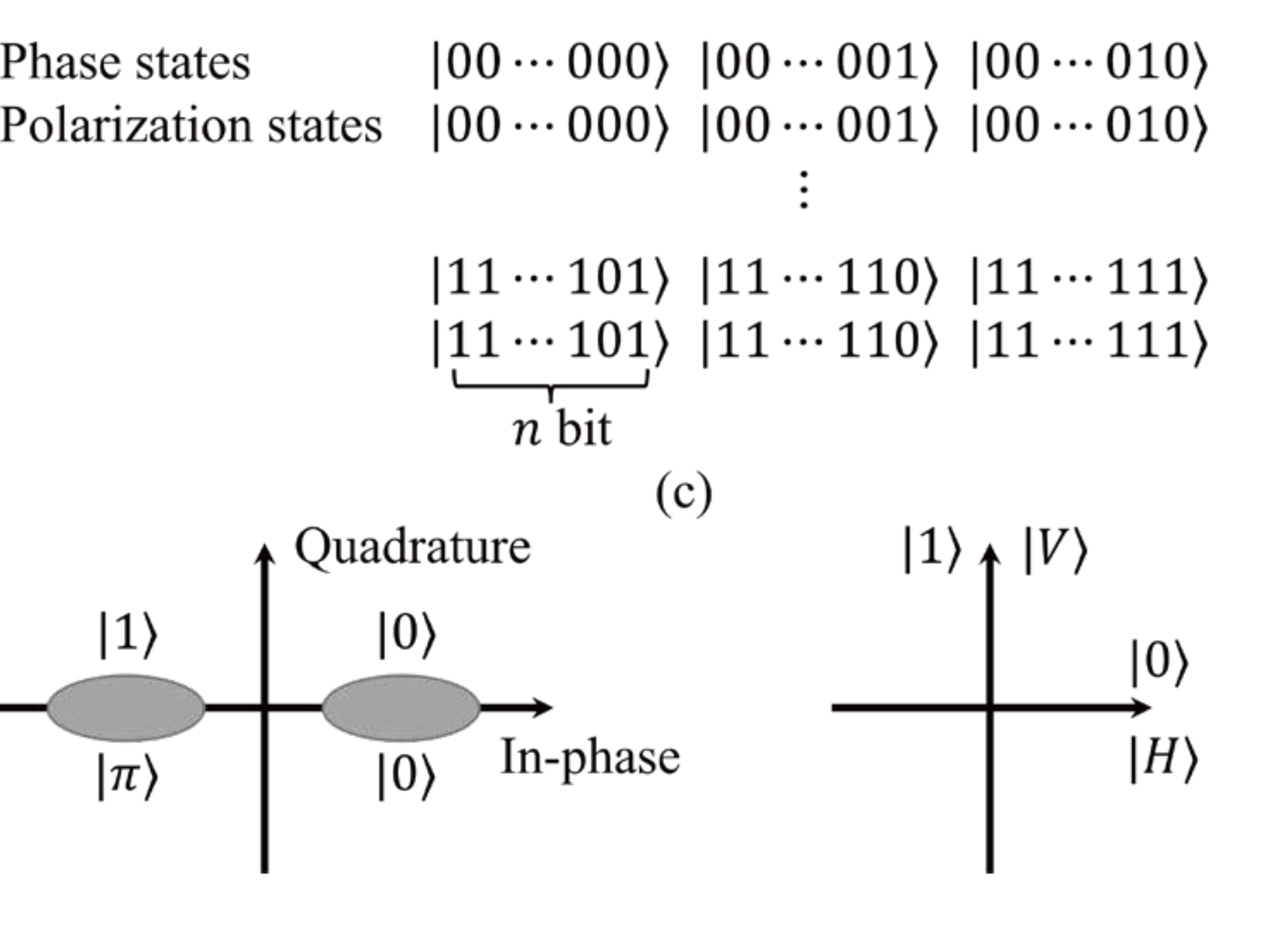}
    
    \caption{(a)The initialized states of $n\cdot2^{n}$ DOPO pulses. Each double line represents one group of DOPO pulses, and all information on the
    starting point of the first word of DVs are encoded by groups of phase
    and polarization states, respectively; (b)(c)the representation relations
    between the polarization and phase of DOPO and represented qubit state.\label{fig:DOPO initialization}}
\end{figure}

We demonstrate an optical-based implementation of the quantum algorithm
by employing a network of degenerate optical parametric oscillators
(DOPOs), as illustrated in Fig.~\ref{fig:optial setup}. Each DOPO
pulse has two degrees of freedom: in-phase ($0$ or $\pi$) and polarization
(horizontal or vertical), shown in Fig.~\ref{fig:DOPO initialization}(b)
and (c), representing a two-qubit eigenstate. Following this construction,
the control $R_{C}$ and work register $R_{W}$ are represented by
the in-phase and polarization of the DOPO pulses train running in
the orange ring optical fiber cavity in Fig.~\ref{fig:optial setup}.
The implementation of the entangled desired state in Eq.~\ref{eq:desired _state}
requires $2^{n}$ groups of $n=16\lambda$ DOPO pulses and each group
of pulses represents a coupled eigenstate $|\zeta,\zeta\rangle$.
Totally $n\cdot2^{n}$ number of DOPO pulses are required in the cyclic
fiber, and the states in the cavity are shown in Fig.~\ref{fig:DOPO initialization}(a).
The time-multiplexed DOPO network is formed by the pump-powered PSA
in the single fiber-ring containing pulses with the round trip time
$T_{rt}$.

A $1536$ nm pulsed laser source is used to generate pulses and power
the DOPO in the ring cavity. The beam combiner (BC) 1 splits the source
into two paths, one of which behaves as input pulses of the machine
and passes through a second-harmonic generation (SHG) process by a
PPLN. The other is split again by BC3 to serve as a local oscillator
(LO) for balanced homodyne detection (BHD) and feedback pulses after
being modulated by phase and intensity modulators controlled by FPGA\ \cite{li2022scalable},
shown as the blue control lines in Fig.~\ref{fig:DOPO initialization}.
The input signal after SHG is used to synchronously pump the PPLN
for a PSA in the fiber ring. All the DOPO pulses are generated with
horizontal polarization initially, due to the type $0$ phase matching
proposition of the PSA process\ \cite{andrekson2020fiber}. The optical
switch, which is realized by an EOS and a FPC, is connected for allowing
horizontally polarized pulses to pass through initially. The EOS is
initially switched to disconnected with the FPC and controlled by
the FPGA through the magenta control line in Fig.~\ref{fig:DOPO initialization}.
A fiber stretcher is connected for stabilizing the state of pulses
and maintaining the length of the circulated cavity. In other words,
the cavity length compensates for the relative phase difference, fixed
at $0$ or $\pi$, between the signal pulse and pump pulse, which
are maintained synchronously\ \cite{li2022scalable}.

The initialization of the machine is achieved by horizontally polarized
and in-phase $0$ DOPO pulses representing the initial $|0\rangle$
state. To implement the iterations of $DP$ gating and achieve the
entanglement, the FPGA controls the IM and PM to adjust the feedback
pulses to tune the cavity state into the machine state in Eq.~\ref{eq:approx_state_after_DP}.
The tail state $|\psi\rangle$ is neglected because we selectively
tune the phase and polarization of each pulse in a group to represent
the entangled desired state directly. The polarization of each pulse
is recorded in the FPGA and will not be physically changed before
performing a measurement of the machine. Then, a BHD measures the
in-phase component $c_{j}$ of each pulse, which collects the phase
information $\zeta$ of each group of pulses and sends to the FPGA\ \cite{chi2011balanced,li2022scalable}.
The phase part of the DOPO pulse is $0$ or $\pi$ corresponding to
$c_{j}$ is positive or negative, respectively. The FPGA uses the
inputs of combined phase and polarization information $(\zeta,\zeta)$
to control the IM and PM to modulate the corresponding feedback pulses
$f_{i}$ that interfere with the original pulses and tune the in-phase
components of each group of pulses\ \cite{okawachi2020nanophotonic,marandi2012all}.
The state represented by the group of pulses is changed from the eigenstate
$|\zeta,\zeta\rangle$ to the hashed state $|\zeta,f(\zeta)\rangle$,
which implements the DV-validity authenticating gate $U_{f}$.

The Grover search is realized by a polarization filter that retains
the entire group of pulses with pure horizontal polarization. Before
performing a measurement, the FPGA controls the EOS to adjust the
polarization of groups of DOPO into their corresponding states $|f(\zeta)\rangle$.
The machine state shown in Eq.\ \ref{eq:final state} is finally
presented by the DOPO trains in the optical cavity and sent to the
BHD. As the polarizations of DOPOs are changed before they are measured,
the Grover search then is equivalent to a PBS, where all the states
with any vertical polarization will be broken and with pure horizontal
polarization will be preserved and sent into BHD, which is equivalent
to the Grover search to get the states with $|0\rangle$ in the work
register $R_{W}$.

\section{Conclusions\label{sec:Summary}}

We propose a quantum algorithm for seeding disturbance vectors, which
describe differential paths that would successfully lead to collisions
on SHA-1 hashing results. The algorithm has a query complexity of $\mathcal{O}(2^{n/2+1})$
where $n$ is the number of qubits that encodes the addresses of disturbance
vectors. A simulation of the algorithm is presented, successfully
verifying the algorithmic validity where the collision-inducing vectors
have their probabilities amplified under the algorithmic steps of
phase-diffusion gating and Grover search. We further reduce the search
space by confining the candidate disturbance vectors into a specific
type, within which the quantum algorithm is still valid. In addition,
an optical implementation of our algorithm is proposed using a DOPO
network of coherent laser pulses running in an ring fiber cavity.
These results highlight that the proposed algorithm is imminently
implementable and is able to find the same disturbance vectors with
lower computational complexity than its classical counterparts.

\section*{Reference}

\bibliographystyle{iopart-num}
\bibliography{iopart-num}

\end{document}